\documentclass[12pt,a4paper]{article}
\usepackage{graphicx}

\def\,{\ifmmode\mskip\thinmuskip\else\leavevmode\thinspace\fi}
\def\nn{\nonumber}
\newcommand\unit[1]{\,\textmd{#1}}

\title {Scalar and pseudoscalar meson pole terms in the hadronic
light--by--light contributions to $a_\mu^{had}$}

\author{E.~Barto\v{s}$^1$, A.-Z.~Dubni\v{c}kov\'a$^1$,
S.~Dubni\v{c}ka$^2$, E.~A.~Kuraev$^3$ \\ and E.~Zemlyanaya$^4$}
\date{}

\begin{document}
\maketitle

\begin{center} {
$^{1}$ \it Dept. of Theor. Physics, Comenius Univ., Bratislava,
Slovak Republic \\
$^{2}$ \it Inst. of Physics, Slovak Acad. of Sci., Bratislava,
Slovak Republic \\
$^{3}$ \it Bogoliubov Lab. of Theor. Physics, JINR Dubna, 141980
Dubna, Russia \\
$^{4}$ \it Lab. of Information Technologies, JINR Dubna, 141980
Dubna, Russia}
 \\[.2cm]
\end{center}

\vspace*{0.5cm}

\begin{abstract}

     Third QED order hadronic light--by--light (LBL) contributions\\
$a_\mu^{LBL}(M)$ to the anomalous magnetic moment of the muon
$a_\mu^{had}$ from the pole terms of scalar $\sigma$, $a_0(980)$
and pseudoscalar $\pi^0$, $\eta$, $\eta '$ mesons  $(M)$ in the
framework of the linearized extended Nambu--Jona--Lasinio model
are evaluated. The off--shell structure of the
photon--photon--meson vertices is taken into account by means of
constituent quark triangle loops. The mass of the quark is taken
to be $m_u=m_d=m_q=(280 \pm 20) \unit{MeV}$. The unknown strong
coupling constants of $\pi^0, \eta, \eta '$ and $a_0$ mesons with
quarks are evaluated in a comparison of the corresponding
theoretical two--photon widths calculated in the framework of our
approach with experimental ones. The $\sigma$--meson coupling
constant is taken to be equal to $\pi_0$--meson coupling constant
as it follows from the linearized Nambu--Jona--Lasinio model
Lagrangian. Then one obtains $a_\mu^{LBL}(\pi_0)$=$(81.83 \pm
16.50) \times 10^{-11}$, $a_\mu^{LBL}(\eta)$=$(5.62 \pm 1.25)
\times 10^{-11}$, $a_\mu^{LBL}(\eta ')$=$(8.00 \pm 1.74) \times
10^{-11}$, $a_\mu^{LBL}(\sigma)$=$(11.67 \pm 2.38) \times
10^{-11}$ and $a_\mu^{LBL}(a_0)$=$(0.62 \pm 0.24) \times
10^{-11}$. The total contribution of meson poles in LBL is
$a_\mu^{LBL}(M)$=$(107.74 \pm 16.81) \times 10^{-11}$.

\end{abstract}
\newpage

\section{Introduction}

The muon is described by the Dirac equation and its magnetic
moment is related to the spin by means of the expression
\begin{equation}
  \vec{\mu}=g\left({e\over 2m_\mu}\right)\vec{s}
\end{equation}
where the value of the gyromagnetic ratio $g$ is predicted (in the
absence of the Pauli term) to be exactly 2.

In fact, however, the interactions existing in nature modify $g$
to be exceeding the \mbox{value 2} because of the emission and
absorption of virtual photons (electromagnetic effects),
intermediate vector and Higgs bosons (weak interaction effects)
and the vacuum polarization into virtual hadronic states (strong
interaction effects).

In order to describe this modification of $g$ theoretically, the
magnetic anomaly was introduced by the relation
\begin{eqnarray}
a_\mu&\equiv& \frac{g-2}{2}=\\ \nn
&&a_\mu^{(1)}\left({\alpha\over\pi} \right) +
\left(a_\mu^{(2)QED}+a_\mu^{(2)had}\right)
\left({\alpha\over\pi}\right)^2 + a_\mu^{(2)weak} +
O\left({\alpha\over\pi}\right)^3
\end{eqnarray}
where $\alpha=1/137.03599976(50)$ \cite{Review} is the fine
structure constant.

Here we would like to make a note that from all three charged
leptons ($e^-$, $\mu^-$, $\tau^-$) the muon magnetic anomaly is
the most interesting object for theoretical investigations due to
the following reasons:

\begin{enumerate}
  \item[$\left.i\right)$] it is one of the best measured
   quantities in physics \cite{Brown}
\begin{equation}\label{eq:2}
  a_\mu^{exp}=(116 592 020 \pm 160)\times 10^{-11}
\end{equation}

  \item[$\left.ii\right)$] its accurate theoretical evaluation
  provides an extremely clean test of "Electroweak theory" and may
  give hints on possible deviations from Standard Model (SM)

  \item[$\left.iii\right)$] the new measurement \cite{Carey}
  in BNL is expected to be performed with a
  definitive accuracy
\begin{equation}\label{eq:3}
  \Delta a_\mu^{exp}=\pm 40\times 10^{-11}
\end{equation}
i.e. it is aimed at obtaining a factor 4 in a precision above that
of the last E--821 measurements (\ref{eq:2}).
\end{enumerate}

At the aimed level of the precision (\ref{eq:3}) a sensibility
will already exist to contributions \cite{Czarnecki:1,Degrassi}
\begin{equation}
  a_\mu^{(2,3)weak}=(152 \pm 4)\times 10^{-11},
\end{equation}
arising from single-- and two--loop weak interaction diagrams.

However, as we have mentioned above, the muon magnetic anomaly may
also contain contributions from a new physics, which can be
revealed, we hope, in a comparison of $a_\mu^{exp}$ with an
accurate theoretical evaluation of $a_\mu^{th}$. While QED and
weak interaction contributions to $a_\mu^{th}$ seem to be
estimated very reliably, there is still opened a door for
improvements in hadronic contributions.

As the most critical from all hadronic contributions are the
light--by--light (LBL) meson pole terms \cite{Melnikov}, we
recalculate the third--order hadronic LBL contributions to the
anomalous magnetic moment of the muon $a_\mu^{had}$ from the pole
terms of the scalar $\sigma$, $a_0$ and pseudoscalar $\pi^0$,
$\eta$, $\eta'$ mesons $(M)$ in the framework of the linearized
extended Nambu--Jona--Lasinio model. The reason for the latter are
predictions of series of recent papers
\begin{eqnarray}  \label{eq:60}
a_\mu^{LBL}&=&(-52 \pm 18) \times 10^{-11}\quad[7] \nn\\
a_\mu^{LBL}&=&(-92 \pm 32) \times 10^{-11}\quad[8]  \nn\\
a_\mu^{LBL}&=&(-79.2 \pm 15.4) \times 10^{-11} \quad[9]\\
a_\mu^{LBL}&=&(+83 \pm 12) \times 10^{-11} \quad[10] \nn  \\
a_\mu^{LBL}(\pi_0)&=&(+58 \pm 10)\times 10^{-11} \quad[10, 11] \nn
\end{eqnarray}
which differ not only in the magnitude, but even in the sign.
Moreover, in these papers only the pseudoscalar pole contributions
were considered. In our paper we include the scalar meson
$(\sigma, a_0)$ pole contributions as well.

The crucial point is description of transition form factor
$\gamma^*\rightarrow M\gamma^*$. Current methods used for this aim
are the ChPT and the vector--meson--dominance (VMD) model.

   In this paper the corresponding transition form
factors by the constituent quark triangle loops with colourless
and flavourless quarks with charge equal to the electron one are
represented. (An application of a similar modified constituent
quark triangle loop model for a prediction of the pion
electromagnetic form factor behavior can be found in
\cite{Dubnicka:1}, where also a comparison with the naive VMD
model prediction is carried out.) The mass of the quark in the
triangle loop is always taken to be  $m_u=m_d=m_q= (280 \pm 20)
\unit{MeV}$ \cite{Volkov}, which was determined in the framework
of the chiral quark model of the Nambu--Jona--Lasinio type by
exploiting the experimental values of the pion decay constant, the
$\rho$--meson decay into two--pions constant, the masses of pion
and kaon and the mass difference of $\eta$ and $\eta '$ mesons.
The unknown strong coupling constants of $\pi^0,\eta,\eta '$ and
$a_0$ mesons with quarks are evaluated in a comparison of the
corresponding theoretical two-photon widths with experimental
ones. The $\sigma$--meson coupling constant is taken to be equal
to $\pi^0$--meson coupling constant as it follows from the
corresponding Lagrangian. The $\sigma$--meson mass is taken to be
$m_\sigma$=$(496 \pm 47) \unit{MeV}$  as an average of the values
recently obtained experimentally from the decay $D^+ \rightarrow
\pi^-\pi^+\pi^+$ \cite{Aitala} and excited $\Upsilon$ decay
\cite{Komada} processes.

   As a result we present explicit formulas for $a_\mu^{LBL}(M)$
$(M=\pi^0, \eta, \eta ', \sigma, a_0)$ in terms of Feynman
parametric integrals of 10--dimensional order, which subsequently
are calculated by MIKOR method \cite{Korobov}.

   The paper is organized as follows. In the next section all definitions
and a derivation of basic relations is presented. More detail can
be found in Appendix A and in Appendix B. The last section to
numerical results and discussion is devoted.

\section{Meson pole terms in light--by--light contributions to
$a_\mu^{had}$}

\begin{figure}[ht]
\begin{center}
\includegraphics{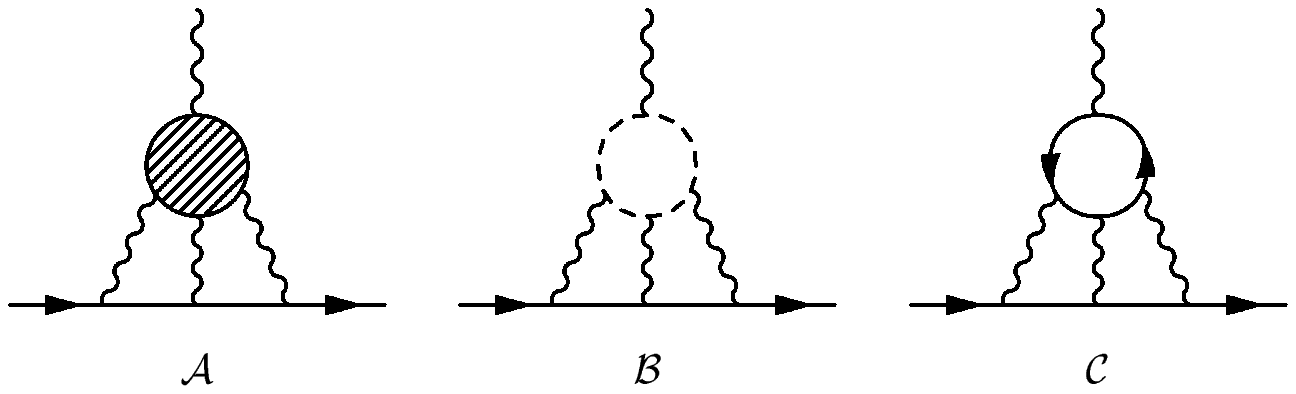}
\caption{Third order hadronic light--by--light scattering
contribution to $a_\mu^{had}$ ($\mathcal{A}$) and class of
pseudoscalar meson square loop diagrams ($\mathcal{B}$) and quark
square loop diagrams ($\mathcal{C}$) contributing to
($\mathcal{A}$).} \label{fig:1}
\end{center}
\end{figure}

\begin{figure}[th]
\begin{center}
\includegraphics{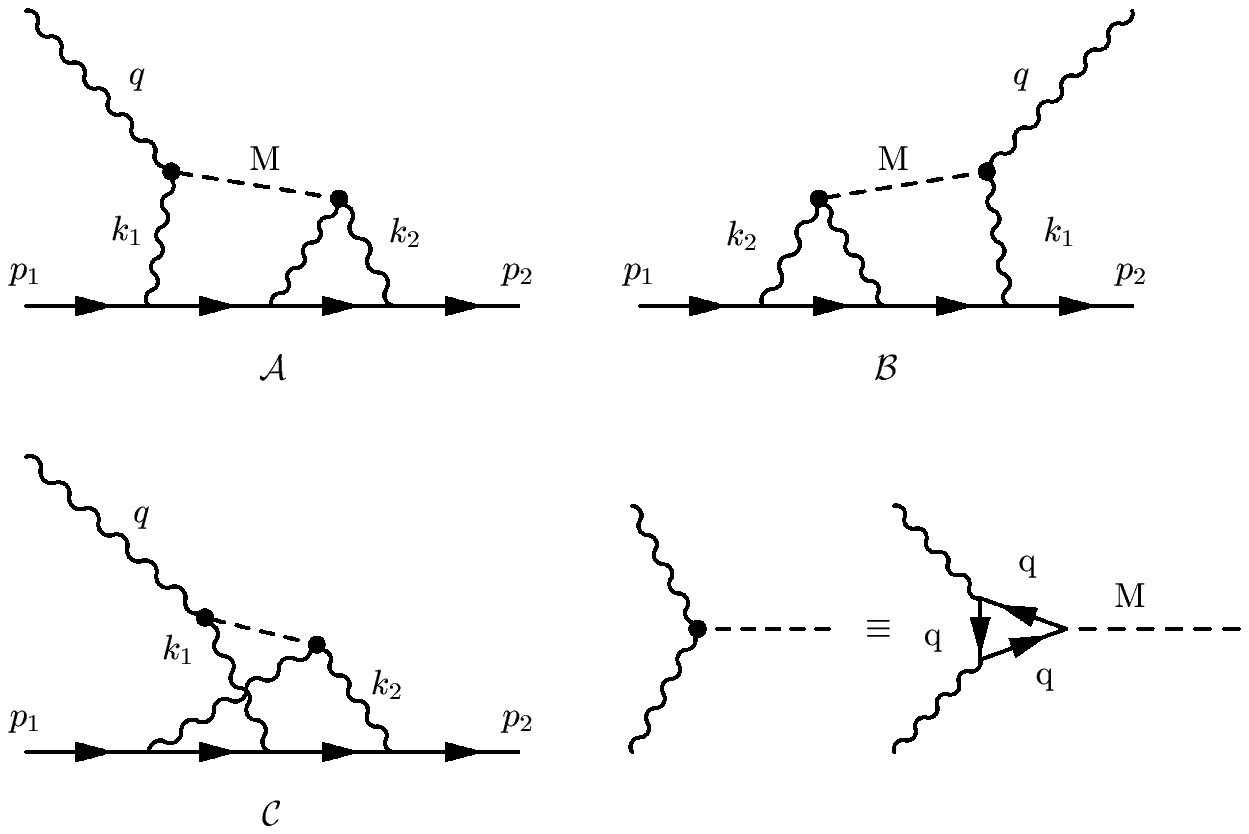}
\caption{Meson (M) pole diagrams in the third order hadronic
light--by--light scattering contributions to $a_\mu^{had}$.}
\label{fig:2}
\end{center}
\end{figure}

The third QED order hadronic LBL scattering contributions to
$a_\mu^{had}$ are generally represented by the diagram in
\mbox{Fig. \ref{fig:1}$\mathcal{A}$} , which contains a class of
pseudoscalar mesons ($\pi^\pm$, $K^\pm$ and also $K_{S,L}^0$)
square loop diagrams (\mbox{Fig. \ref{fig:1}$\mathcal{B}$}), a
class of quark square loop diagrams (\mbox{Fig.
\ref{fig:1}$\mathcal{C}$}) and scalar $\sigma$, $a_0(980)$ and
pseudoscalar $\pi^0$, $\eta$, $\eta '$ meson pole diagrams
(\mbox{Fig. \ref{fig:2}}), where the off-shell structure of the
photon--photon--meson vertices is taken into account by means of
flavourless and colourless constituent quark triangle loops. Here
the interaction of mesons (M) with quarks is described by the
linearized Nambu--Jona--Lasinio type Lagrangian
\begin{equation}
\mathcal{L}_{q\bar{q}M}=g_M\bar{q}(x)\left[\sigma(x)
+i\pi(x)\gamma_5\right]q(x),
\end{equation}
with unknown strong coupling constant $g_M$. The latter for
$\pi^0$, $\eta$, $\eta '$ and $a_0$ mesons are evaluated by a
comparison of the corresponding theoretical two-photon widths with
the experimentally determined values \cite{Review}.

   For the on--mass shell scalar $(S)$ and pseudoscalar $(P)$ mesons
decays one can write the following matrix elements

\begin{eqnarray}
&&M(S(p)\rightarrow\gamma(k_1)+\gamma(k_2))=
-\frac{i\alpha g_S}{\pi m_q}K(m_S/m_q)(k_1k_2g_{\mu\nu}-\\
&&\hspace{5cm} -k_{1\nu}k_{2\mu}) e_{1\nu}(k_1)
e_{2\mu}(k_2),\nn\\
&&M(P(p)\rightarrow\gamma(k_1)+\gamma(k_2))= \frac{\alpha g_P}{\pi
m_q}J(m_P/m_q)(k_1k_2e_1e_2),
\end{eqnarray}
with

\begin{eqnarray}
K(z)=2\int\limits_0^1dx\int\limits_0^{1-x}\frac{1-4xy}{1-xyz^2}dy,\;\;
J(z)=\frac{2}{z^2}\int\limits_0^1\frac{dx}{x}\ln[1-z^2x(1-x)]
\end{eqnarray}
and

$$p^2=m_M^2,\; k_1^2=k_2^2=0,\;
(abcd)=\epsilon_{\alpha\beta\gamma\delta}a^\alpha b^\beta c^\gamma
d^\delta.$$

They lead to the theoretical two--photon widths (for more detail
see Appendix A)
\begin{eqnarray}
\Gamma_S^{\gamma\gamma}=\frac{\alpha^2m_S^3g_S^2}{64\pi^3 m_q^2}
K^2({m_S/m_q}), \\
\Gamma_{P}^{\gamma\gamma}=\frac{\alpha^2m_P^3g_P^2}{64\pi^3
m_q^2}J^2({m_P/m_q}).
\end{eqnarray}
dependent on the scalar $g_S$ and pseudoscalar $g_P$ meson
coupling constants with quarks, respectively.

   Taking the meson masses with errors and the following world averaged
values of the two--photon widths from the last Review of Particle
Physics \cite{Review}, $\Gamma_{\pi^0}^{\gamma\gamma}=(0.008 \pm
0) \unit{keV}$, $\Gamma_\eta^{\gamma\gamma}=(0.464 \pm 0.044)
\unit{keV}$, $\Gamma_{\eta '}^{\gamma\gamma}=(4.282 \pm 0.339)
\unit{keV}$ and $\Gamma_{a_0}^{\gamma\gamma}=(0.24 \pm 0.08)
\unit{keV}$, the quark mass value $m_q=(280 \pm 20) \unit{MeV}$,
one finds the meson coupling constants with quarks as follows
\begin{eqnarray}
g_{\pi^0}^2=9.120 \pm 1.305, g_\eta^2=2.220 \pm 0.378, \\
g_{\eta '}^2=6.708 \pm 1.096, g_{a_0}^2=0.757 \pm 0.277. \nn
\end{eqnarray}

   Since there is no experimental result on the $\sigma$-meson decay
into two--photons up to now, we identify $g_\sigma^2$ with
$g_{\pi^0}^2$ as it follows from the Lagrangian (7).

   Now we are ready to evaluate LBL meson pole terms
contributions to $a_\mu^{had}$.

   Usually the fermion--photon vertex function is written in the form
\begin{eqnarray}
&\Gamma_\mu=\bar{u}(p_2)\hat{V}_\mu(q)u(p_1),& \\
&\hat{V}_\mu(q)=\gamma_\mu F_1(q^2)+\frac{1}{4m}[\hat{q},
\gamma_\mu]F_2(q^2),\quad q=p_2-p_1,& \nn
\end{eqnarray}
with on mass--shell muons, i.e. $p_1^2=p_2^2=m^2$ ($m$ is the muon
mass). Then the expression for the muon anomalous magnetic moment
is
\begin{eqnarray}
a_\mu&=&\frac{(g-2)_\mu}{2}=F_2(0)\\\nn
&=&-\frac{3}{16m^2}\lim_{q^2\to 0}\frac{1}
{q^2}Sp(\hat{p}_2+m)\hat{V}_\mu(q)(\hat{p}_1+m)\mathcal{P}_\mu(q),
\end{eqnarray}
with a projection operator
$$\mathcal{P}_\mu(q)=q^2\gamma_\mu+{m\over 3}[\hat{q},\gamma_\mu].$$

The gauge invariant set of Feynman diagrams containing the LBL
scattering block with meson pole intermediate states is drawn in
Fig.~2.

Their contributions to $a_\mu^{had}$ are
\begin{eqnarray}
&a_\mu^{LBL}(S)+a_\mu^{LBL}(P)&=\\ \nn
&&-\frac{3}{16m^2}\lim_{q^2\to 0}\frac{1}
{q^2}Sp(\hat{p}_2+m)\hat{V}_\mu^{S+P}(\hat{p}_1+m)\mathcal{P}_\mu(q),
\end{eqnarray}
with
\begin{eqnarray}
\hat{V}_\mu^{S+P}=\frac{\alpha}{2^6\pi^3}\int\frac{d^4k_1 d^4k_2}
{(i\pi^2)^2}O^{\nu\lambda\sigma}\left[\frac{T^{(P)}_{\mu\nu\lambda\sigma}}
{k_1^2-m_P^2}+\frac{T^{(S)}_{\mu\nu\lambda\sigma}}{k_1^2-m_S^2}\right]
\frac{1}{k_1^2k_2^2(k_1-k_2)^2}, \\
O=2O_1+O_2,\; \nn \\
O_1^{\nu\lambda\sigma}=\gamma^\sigma\frac{\hat{p}_2-
\hat{k}_2+m}{(p_2-k_2)^2-m^2}\gamma^\lambda\frac{\hat{p}_1-
\hat{k}_1+m}{(p_1-k_1)^2-m^2}\gamma^\nu, \nn\\
O_2^{\nu\lambda\sigma}=\gamma^\sigma\frac{\hat{p}_2-
\hat{k}_2+m}{(p_2-k_2)^2-m^2}\gamma^\nu\frac{\hat{p}_1+
\hat{k}_1-\hat{k}_2+m}{(p_1+k_1-k_2)^2-m^2}\gamma^\lambda \nn
\end{eqnarray}
where
\begin{equation}
T^{(P)}_{\mu\nu\lambda\sigma}=\frac{4\alpha^2g^2_P}{\pi^2m_q^2}
(\mu\nu k_1q)(\lambda\sigma k_2k_1)f_1(k_1^2)f_2(k_1,k_2),
\end{equation}
with
\begin{eqnarray}
f_1(k_1^2)=\int\limits_0^1\frac{m_q^2}{m_q^2-x(1-x)k_1^2}dx,
~f_2(k_1,k_2)=2\int\limits_0^1 dx_1\int\limits_0^{1-x_1}
\frac{m_q^2}{d}dx_2,\\
d=m_q^2-k_2^2x_3(1-x_3)-k_1^2x_2(1-x_2)+2k_1k_2x_2x_3, \nn \\
x_3=1-x_1-x_2. \nn
\end{eqnarray}
Here we would like to note that the functions $f_1(k_1^2)$,
$f_2(k_1,k_2)$ at the limit of an infinite value of the quark
mass, $m_q\to\infty$, tend to one.

   The expression for $T^{(S)}$ has a similar form
\begin{eqnarray}
&&T^{(S)}_{\mu\nu\lambda\sigma}=-\frac{4\alpha^2g_S^2}
{\pi^2m_q^2}(g_{\mu\nu}k_1q-
k_{1\mu} q_\nu)\tilde{f}_1(k_1^2)f_{\lambda\sigma}(k_1,k_2), \\
&&\qquad\tilde{f}_1(k_1^2)=\int\limits_0^1\frac{m_q^2(1-2x(1-x))dx}
{m_q^2-x(1-x)k_1^2}, \nn\\
&&\qquad f_{\lambda\sigma}(k_1,k_2)=2\int\limits_0^1dx_1\int
\limits_0^{1-x_1}\frac{m_q^2}{d}\left\{
(1-4x_1x_2)\left[g_{\lambda\sigma}k_2(k_1-k_2)-\right.\right.\nn\\
&&\hspace{2cm}\left.\left.k_{2\lambda}(k_1-k_2)_\sigma\right]+
2(k_1-k_2)^2x_1(2x_1-1)g_{\lambda\sigma}\right\}dx_2. \nn
\end{eqnarray}

   Usual procedure of performing the loop--momenta integration using
the Feynman parameters method (see Appendix B) leads to
\begin{equation}
a_\mu^{LBL}(M)=-R_MF_M(m_M,m_q)
\end{equation}
where
\begin{equation}
R_M=\frac{3m^2\alpha^2g_M^2}{64\pi^5m_q^2}
\end{equation}
and
\begin{eqnarray}
&F_P(m_P,m_q)=2M_a+M_b,\quad F_S(m_S,m_q)=2N_a+N_b.&\nn
\end{eqnarray}
The $M_i,N_i$ are given by the expressions
\begin{eqnarray}
M_a=\int d\Gamma_a\frac{1}{zD_1^2}
\left[\frac{A^P_a}{D_1}+B^P_a\right]; \nn \\  M_b=\int
d\Gamma_b\frac{1}{\sigma^2D_2^2}
\left[\frac{A^P_b}{D_2}+B^P_b\right]; \\  N_a=\int
d\Gamma_a\frac{1}{zD_1^2} \left[\frac{A^S_a}{D_1}+B^S_a\right]; \nn \\
N_b=\int d\Gamma_b\frac{1}{\sigma^2D_2^2}
\left[\frac{A^S_b}{D_2}+B^S_b\right],\nn
\end{eqnarray}
with $d\Gamma_{a,b}$ the 10--dimensional integrals on Feynman
parameters
\begin{eqnarray}
&&\quad\int d\Gamma_a=\tau^4\int\limits_0^1\frac{dx}{x\bar{x}}\int
\limits_0^1dx_1\int\limits_0^{\bar{x}_1}\frac{dx_3}{x_3\bar{x_3}}
\int\limits_0^1dz_1\int\limits_0^{\bar{z}_1}d z_2\int
\limits_0^1\alpha_2\alpha_3^2d^3\alpha dy\rho\bar{\rho}^2d\rho, \\
&&\quad\nn \int
d\Gamma_b=\tau^4\int\limits_0^1\frac{dx}{x\bar{x}}\int
\limits_0^1dx_1\int\limits_0^{\bar{x}_1}\frac{dx_3}{x_3\bar{x_3}}
\int\limits_0^1 dz_1\int\limits_0^{\bar{z}_1}dz_2\int\limits_0^1
\alpha_2\alpha_3^2\alpha_4^3 d^4\alpha\rho\bar{\rho}^2d\rho, \\
&&\nn \bar{x}=1-x, \bar{z_1}=1-z_1, \tau=\frac{m_q}{m},
d^3\alpha=d\alpha_1d\alpha_2d\alpha_3,
d^4\alpha=d\alpha_1d\alpha_2d\alpha_3d\alpha_4.
\end{eqnarray}
The explicit expressions for polynomials $A$, $B$ are
\begin{eqnarray}
A^P_a&=&\frac{8y\rho}{3z}\left[2a_1^2+2r_1a_1(a_1+b_1)-
r_1^2a_1b_1\right]-8r_1^3, \nn \\
B^P_a&=&4+12r_1+4a_1b_1\frac{y\rho}{z}, \nn \\
A^P_b&=&\frac{8\rho}{3\sigma}[a_2^3-r_2^2a_2b_2(b_2-1)]-
\frac{16}{3}r_2^2a_2+\frac{8}{3}r_2^3(1-2b_2), \nn \\
B^P_a&=&4a_2-\frac{8}{3}+r_2(8b_2-4)+\frac{4\rho}
{\sigma}a_2b_2(b_2-1),\nn \\
A^S_a&=&\frac{y\rho}{z}\left[-8a_1^3-16r_1a_1(a_1b_1-
\frac{1}{3}a_1+\frac{1}{3}b_1)\right. \nn\\
&-&\left.8r_1^2(a_1b_1-\frac{2}{3}a_1+\frac{2}{3}b_1-
\frac{2}{3})b_1\right]d_1(x_1,x_2) \nn\\
&+&\frac{y\rho}{z}\left[(16/3)a_1^2(a_1+1)+\frac{32}{3}r_1(b_1-1)
(a_1+1)a_1\right.\\
&+&\left.r_1^2\frac{16}{3}(a_1+1)(b_1-1)^2\right]d_2(x_1), \nn\\
B^S_a&=&(8a_1+\frac{16}{3})d_2(x_1)-(4+12a_1)d_1(x_1,x_2) \nn\\
&-&\frac{8y\rho}{z}(a_1+1)(b_1-1)^2d_2(x_1) \nn\\
&+&\frac{y\rho}{z}(12a_1b_1-8a_1+8b_1-8)b_1d_1(x_1,x_2),\nn\\
A_b^S&=&\frac{\rho}{\sigma}\left\{\left[\frac{16}{3}
a_2^2(a_2-1)+\frac{32}{3}r_2a_2(a_2-1)(b_2-1)\right.\right.\nn\\
&+&\left.\frac{16}{3}r_2(a_2-1)(b_2-1)^2\right]d_2(x_1)\nn\\
&+&\left.\left[-\frac{16}{3}a_2^2+\frac{16}{3}r_2^2b_2(b_2-1)
\right]d_1(x_1,x_2)\right\}, \nn\\
B_b^S&=&\left[\frac{8}{3}-\frac{8\rho}{\sigma}
b_2(b_2-1)\right]d_1(x_1,x_2)\nn\\
&+&\left[8a_2-\frac{16}{3}-\frac{8\rho}
{\sigma}(a_2-1)(b_2-1)^2\right]d_2(x_1)\nn
\end{eqnarray}
with
\begin{eqnarray}
&&a_1=\alpha_3\bar{\alpha}_2,\; b_1=\alpha_2\alpha_3\gamma_1,
a_2=\alpha_4(\alpha_3\bar{\alpha}_2+\bar{\alpha}_3),\;\\
&&b_2=\alpha_4(\alpha_3\gamma_1+\bar{\alpha}_3),\;
r_1=\rho(\bar{y}-c_1y),\; r_2=-c_2\rho,\;\nn\\
&&d_1(x_1,x_2)=1-4x_1x_2,\; d_2(x_1)=2x_1(2x_1-1).\nn
\end{eqnarray}
Quantities $c_{1,2}$, $\gamma_1$, $z$, $D_1$, $D_2$ and $\sigma$
are given in Appendix B. \vspace{3cm}

\section{Results and discussion}

The  result for the meson pole contributions of LBL type has the
form
\begin{equation}
a_\mu^{LBL}(M)=a_\mu^{LBL}(\pi^0)+a_\mu^{LBL}(\eta)+a_\mu^{LBL}(\eta')+
a_\mu^{LBL}(\sigma)+a_\mu^{LBL}(a_0)
\end{equation}
where
\begin{eqnarray}
a_\mu^{LBL}(\pi^0) &=& -R_{\pi^0}F_{\pi^0}(m_{\pi^0}, m_q) \nn \\
a_\mu^{LBL}(\eta) &=& -R_{\eta}F_{\eta}(m_\eta, m_q) \nn \\
a_\mu^{LBL}(\eta ') &=& -R_{\eta '}F_{\eta '}(m_{\eta '}, m_q)\\
a_\mu^{LBL}(\sigma) &=& -R_{\sigma}F_{\sigma}(m_{\sigma}, m_q) \nn \\
a_\mu^{LBL}(a_0) &=& -R_{a_0}F_{a_0}(m_{a_0}, m_q). \nn
\end{eqnarray}

Now, taking the values (13) of $g_M^2$ determined in Section 2,
the mass of the quark $m_q=(280 \pm 20)\unit{MeV}$ \cite{Volkov},
the mass of the $\sigma$--meson $m_\sigma=(496 \pm 47) \unit{MeV}$
\cite{Aitala,Komada}, the mass of the muon and the corresponding
scalar and pseudoscalar mesons from the Review of Particle Physics
\cite{Review} one finds
\begin{eqnarray}
a_\mu^{LBL}(\pi^0) &=& (81.83 \pm 16.50) \times 10^{-11} \nn \\
a_\mu^{LBL}(\eta) &=& (5.62 \pm 1.25) \times 10^{-11} \nn \\
a_\mu^{LBL}(\eta ') &=& (8.00 \pm 1.74) \times 10^{-11} \\
a_\mu^{LBL}(\sigma) &=& (11.67 \pm 2.38) \times 10^{-11} \nn \\
a_\mu^{LBL}(a_0) &=& (0.62 \pm 0.24) \times 10^{-11}. \nn
\end{eqnarray}
   Here we would like to note that our value for the $\pi^0$ is in the
framework of error bars consistent with \cite{Knecht,Czarnecki:2}.

   So, the total contribution of meson poles in LBL is
\begin{equation}
             a_\mu^{LBL}(M) = (107.74 \pm 16.81) \times 10^{-11},
\end{equation}
where the resultant error is the addition in quadrature of all
partial errors of (30).

   Together with the contributions of the pseudoscalar meson $(\pi^\pm,
K^\pm, K_{S,L}^0)$ square loops and constituent quark square loops
\cite{Hayakawa:1,Bijnens} it gives
\begin{equation}
            a_\mu^{LBL}(total) = (111.20 \pm 16.81) \times 10^{-11}.
\end{equation}

   The others 3-loop hadronic contributions derived from the hadronic
vacuum polarizations $(VP)$ were most recently evaluated by Krause
\cite{Krause}
\begin{equation}
              a_\mu^{(3)VP} = (-101 \pm 6) \times 10^{-11}.
\end{equation}
Then the total 3-loop hadronic correction from (32) and (33) is
\begin{equation}
       a_\mu^{(3)had} = a_\mu^{LBL}(total) + a_\mu^{(3)VP} =
                        (10.20 \pm 17.28) \times 10^{-11}
\end{equation}
where the errors have been again added in quadratures.

   If we take into account the most recent evaluation \cite{Narison}
of the lowest--order hadronic vacuum--polarization contribution to
the anomalous magnetic moment of the muon
\begin{equation}
              a_\mu^{(2)had} = (7021 \pm 76) \times 10^{-11}
\end{equation}
(it is interesting to notice that this result almost coincides in
the central value and in the error as well with the result
\cite{Dubnicka:2} of one of the authors (S.D.) of this paper
obtained 12 years ago) the pure QED contribution up to 8th order
\cite{Kinoshita}
\begin{equation}
              a_\mu^{QED} = (116 584 705.7 \pm 2.9) \times 10^{-11}
\end{equation}
and the single-- and two--loop weak interaction contribution \cite
{Degrassi}, finally one gets the SM theoretical prediction of the
muon anomalous magnetic moment value to be
\begin{equation}
              a_\mu^{th} = (116 591 888.9 \pm 78.1) \times
              10^{-11}.
\end{equation}

   Comparing this theoretical result with experimental one (3) one
finds
\begin{equation}
              a_\mu^{exp} - a_\mu^{th} = (131 \pm 178) \times 10^{-11}
\end{equation}
which implies very good consistency of the SM prediction for the
anomalous magnetic moment of the muon with experiment.

\section*{Appendix A}
\renewcommand{\theequation}{A.\arabic{equation}}
\setcounter{equation}{0} We put here the details of calculations
of the two--photon widths in the framework of the linearized
Nambu--Jona--Lasinio model.

   The decay matrix element of a pseudoscalar meson $P$ into two
photons has a form (we take into account two directions of fermion
line in the triangle fermion loop)
\begin{equation}
M(P(p)\to\gamma(e_{1\lambda}(k_1))\gamma(e_{2\sigma}(k_2))=
\frac{2i\alpha g_P m_q}{\pi}\int\frac{d^4k}{i\pi^2}\frac{S_1}
{a_1a_2a_3},
\end{equation}
with
\begin{eqnarray}
S_1=\frac{1}{4m_q}Tr(\hat{k}-\hat{k}_2+m_q)\hat{e}_2(\hat{k}+m_q)\hat{e}_1
(\hat{k}+\hat{k}_1+m_q)\gamma_5=i(k_2k_1e_1e_2)\\
a_1=(k+k_1)^2-m_q^2,\; a_2=(k_1-k_2)^2-m_q^2,\;
a_3=k^2-m_q^2,\nn\\\nn
\end{eqnarray}
where we use the definitions
$\gamma_5=i\gamma^0\gamma^1\gamma^2\gamma^3$,
$\frac{1}{4}Sp\gamma_\mu\gamma_\nu\gamma_\lambda\gamma_\sigma=i
\varepsilon_{\mu\nu\lambda\sigma}$, $\varepsilon_{0123}=+1$.

Joining the denominators and performing the loop momentum
integration one obtains
\begin{equation}
M_{P}^{\gamma^*\gamma^*}=-\frac{2g_P\alpha}{\pi m_q}\int
\frac{d^3x\delta(\sum x-1)m_q^2(k_2k_1e_1e_2)} {m_q^2-x_1x_2
p^2-x_1x_3k_2^2-x_2x_3k_1^2}.
\end{equation}
This expression is used for the construction of $f_2(k_1,k_2)$ in
(19).

   For the case of real particles we have
\begin{equation}
M^{\gamma\gamma}_{P}=\frac{\alpha g_P}{\pi m_q}(k_1k_2e_1 e_2)
J(m_P/m_q)
\end{equation}
with $J(z)$ given by (10).

   Note that for the case of $\eta'$--meson the matrix element becomes
complex.
   Numerically one obtains
\begin{equation}
J^2(m_{\pi^0}/m_q)=1.04,\; J^2(m_\eta/m_q)=3.73,\;
\Big|J(m_{\eta'}/m_q)\Big|^2=2.12 \;.
\end{equation}

   For the matrix element of a decay of the scalar meson $S$ into two
photons one can write
\begin{equation}
M(S(p)\to\gamma(k_1,e_1)\gamma(k_2,e_2))=e_{1\mu}e_{2\nu}M^{\mu\nu},
\end{equation}
where
\begin{eqnarray}
&&\qquad M^{\mu\nu}=\frac{2i\alpha
g_Sm_q}{\pi}\int\frac{d^4k}{i\pi^2} \frac{T^{\mu\nu}}{a_1a_2a_3},
\\ \nn &&T^{\mu\nu}=\frac{1}{4m_q}Sp(\hat{k}-\hat{k}_2+m_q)
\gamma_\nu(\hat{k}+m_q)\gamma_\mu(\hat{k}+\hat{k}_2+m_q).
\end{eqnarray}
This quantity suffers from the ultraviolet divergences. It is the
gauge invariance requirement which provides the regularization.
General form of $M^{\mu\nu}(k_1,k_2)$, which satisfies the current
conservation condition
\begin{equation}
M^{\mu\nu}(k_1,k_2)[k_{1\mu};k_{2\nu}]=0,
\end{equation}
has the form
\begin{eqnarray}
M^{\mu\nu}(k_1,k_2)&=&A(g^{\mu\nu}k_1k_2-k_1^\nu k_2^\mu)\\ \nn
&&+ B(k_1^2k_2^\nu-k_1k_2 k_1^\nu)(k_2^2k_1^\mu-k_1k_2k_2^\mu).
\end{eqnarray}
Usual procedure of joining the denominators, the calculation of
the trace, a performation of the loop momentum integration and an
application of the gauge conditions to the muon block, leads to a
final result.

For the case of real particles the structure $B$ disappears and we
have
\begin{equation}
M^{(S\to\gamma\gamma)}=e_1^\mu(k_1)e_2^\nu(k_2)(g_{\mu\nu}k_1k_2-
k_1^\nu k_2^\mu)\frac{-i\alpha g_S}{\pi m_q}K(m_S/m_q),
\end{equation}
which leads to the expression (11) for the width with $K(z)$ given
by (10). Numerically $$|K(m_{a^0}/m_q)|^2= 0.96.$$

\section*{Appendix B}

\renewcommand{\theequation}{B.\arabic{equation}}
\setcounter{equation}{0}

For a performance of the loop momenta integration we use Feynman
trick of "joining" the denominators
\begin{equation}
\frac{1}{a^mb^n}=\frac{(n+m-1)!}{(n-1)!(m-1)!}\int\limits_0^1
\frac{x^{m-1}(1-x)^{n-1}dx}{(ax+b(1-x))^{m+n}}
\end{equation}
and the relation
\begin{equation}
\int\frac{d^4k}{i\pi^2}\frac{(k^2)^m}{(k^2-d)^n}=(-1)^{m+n}
\frac{(m+1)!(n-m-3)!}{(n-1)!d^{n-m-2}},n>m+2.
\end{equation}

For this aim we present the denominators containing $k_2$ in the
form
\begin{equation}
\frac{1}{-x_3\bar{x}_3}\frac{1}{\{1\}\{2\}\{3\}\{4\}}
\left[\frac{1}{\{5\}}; \frac{1}{\{6\}}\right], \nn
\end{equation}
where
\begin{eqnarray}
\{1\}=k_2^2,\{2\}=k_2^2-2p_2k_2,\{3\}=(k_2-k_1)^2,
\{4\}=d/(-x_3\bar{x}_3) \nn\\
\{5\}=k_1^2-2p_1k_1; \{6\}=(p_1+k_1-k_2)^2-m^2,\bar{x}=1-x. \nn
\end{eqnarray}

   Using Feynman parameter $\alpha_1$ to join the denominators $\{3\},\{4\}$
and the parameter $\alpha_2$ for joining the result with $\{2\}$
we have
\begin{eqnarray}
\alpha_2[\alpha_1\{4\}+\bar{\alpha}_1\{3\}]+\bar{\alpha}_2\{2\}=
k_2^2-2k_2a+\alpha_2\Delta_1,a=\alpha_2\gamma_1k_1+\bar{\alpha}_2p_2,\\
\Delta_1=\alpha_2\alpha_1\frac{x_2\bar{x}_2k_1^2-m_q^2}{x_3\bar{x}_3}+
\alpha_2\bar{\alpha}_2k_1^2,\;\gamma_1=\bar{\alpha}_1+
\alpha_1\frac{x_2}{\bar{x}_3}, \bar{a}=1=a. \nn
\end{eqnarray}

   For the diagram in Fig.~2$\mathcal{A}$ we use the parameter $\alpha_3$
for the joining of the denominator $\{1\}$ with the above result
\begin{eqnarray}
&&\hspace{1cm}(k_2-b)^2-zd_1,\\
&&b=\alpha_2 a,
z=(\alpha_2\alpha_3\gamma_1)^2-\alpha_2\alpha_3(\bar{\alpha}_1+
\frac{\alpha_1x_2\bar{x}_2}{x_3\bar{x}_3}), \nn \\
&&d_1=k_1^2+2c_1k_1p_1+\Delta_2, \nn \\
&& c_1=\frac{\alpha_2\alpha_3^2\bar{\alpha}_2\gamma_1}{z},
\Delta_2=\frac{1}{z}\left[m^2\alpha_3^2\bar{\alpha}_2^2+m_q^2
\frac{\alpha_1\alpha_2\alpha_3}{x_3\bar{x}_3}\right].\nn
\end{eqnarray}

   For the diagram in  Fig.~2$\mathcal{B}$ we use the parameter $\alpha_3$
for the joining of the denominator (B.3) with $\{6\}$ and
parameter $\alpha_4$ to join $\{1\}$ with the result
\begin{equation}
(k_2-b_2)^2-\sigma d_2
\end{equation}
where
\begin{eqnarray}
&&\sigma=\alpha_4^2(\alpha_3\gamma_1+\bar{\alpha}_3)^2-
\alpha_4\bar{\alpha}_3-\alpha_2\alpha_3\alpha_4\left(\bar{\alpha}_1+
\alpha_1\frac{x_2\bar{x}_2}{x_3\bar{x}_3}\right),\nn \\
&&b_2=\alpha_4\left[(\bar{\alpha}_3+\alpha_1\gamma_1)k_1+
(\bar{\alpha}_3+\alpha_3\bar{\alpha}_2)p_1\right],\nn \\
&&d_2=k_1^2+2c_2p_1k_1+\delta, \nn\\
&&c_2=\frac{1}{\sigma}\alpha_4\left[\alpha_4(\alpha_3\gamma_1+\
\bar{\alpha}_3)(\alpha_3\bar{\alpha}_2+
\bar{\alpha}_3)-\bar{\alpha}_3\right], \nn\\
&&\delta=\frac{1}{\sigma}\left[m^2\alpha_4^2(\alpha_3\bar{\alpha}_2+
\bar{\alpha}_3)^2+\frac{\alpha_1\alpha_2\alpha_3\alpha_4}
{x_3\bar{x}_3}m_q^2\right]. \nn
\end{eqnarray}

   To perform the $k_1$ integration we use the parameters $z_1,z_2$
to join the denominators containing $k_1^2$
\begin{equation}
\frac{1}{x\bar{x}}\frac{1}{k_1^2(k_1^2-M^2)(k_1^2-M_1^2)}=
\frac{2}{x\bar{x}}\int\limits_0^1dz_1\int\limits_0^{\bar{z}_1}
\frac{dz_2}{d_3^3},
\end{equation}
 with
$$
d_3=k_1^2-z_1M_1^2-z_2M^2,\;M_1^2=\frac{m_q^2}{x\bar{x}}.
$$

For the diagram in Fig.~2 $\mathcal{B}$ we use parameter $\rho$ to
join the denominators $d_2,d_3$
\begin{eqnarray}
&&\hspace{1cm}\rho d_2+\bar{\rho}d_3=(k_1-q_2)^2-m^2D_2;\; \\
&&q_2=-\rho c_2p_1,\;
m^2D_2=m^2(c_2\rho)^2-\delta\rho+\bar{\rho}(z_1M_1^2+z_2M^2). \nn
\end{eqnarray}

   For the diagram in Fig.~2 $\mathcal{A}$ we use parameter $y$ to
join $d_1$ with $\{5\}$
\begin{equation}
yd_1+\bar{y}(5)=k_1^2-2p_1k_1(\bar{y}-c_1y)+y\Delta_1
\end{equation}
and parameter $\rho$ to join the result with $d_3$. As a result
one gets
\begin{eqnarray}
&&(k_1-q_1)^2-D_1m^2;\; q_1=\rho(\bar{y}-c_1y)p_1, \\
&&m^2D_1=m^2\rho^2(\bar{y}-yc_1)^2+\bar{\rho}(z_1M_1^2+z_2M^2)-
\rho y\Delta_1. \nn
\end{eqnarray}

\section*{Acknowledgments}
We are grateful to participants of BLTP Seminar for discussions.
E.~A.~K is grateful to RFBR 01--02--17437 as well as NNC, JINR for
financial support via INTAS Grant No. 97--30494 and to SR grant
2000.

The work was in part also supported by Slovak Grant Agency for
Sciences, Grant No. 2/5085/20 (S.~D.) and Grant No. 1/7068/20
(A.~Z.~D).

\end{document}